# Multispectral narrowband frustrated total internal reflection filter with inclusions of plasmonic nanoparticles

N I Petrov

Scientific and Technological Centre for Unique Instrumentation of Russian Academy of Sciences, 15 Butlerova str., Moscow 117342, Russia

E-mail: petrovni@mail.ru

**Abstract**

A spatial-frequency thin-film filter with inclusions of nanoparticles operating in the visible range of the spectrum is investigated. The effect of nanoparticles embedded in the central and lateral layers of the frustrated total internal reflection filter on the spectral characteristics, taking into account the frequency dispersion, is investigated. It is shown that plasmonic effects cause splitting of the filter bandwidth into a set of narrow-band spectral lines and angular splitting of the incident beam into a set of output beams.

## 1. Introduction

Resonance effects occur when light tunnels through a barrier system due to wave interference. Tunneling effects, which are an optical analogy of quantum mechanical tunneling, have found application in various fields of applied physics, including fiber and integral optics, as well as frustrated total internal reflection spectroscopy. There are various types of devices, such as beam splitters, filters and polarizers, based on interference of light and the frustrated total internal reflection (FTIR) effect in thin films.

The FTIR filter was first proposed in 1947 by Leurgens and Turner [1]. This is a device that uses resonant tunneling of light through a planar dielectric optical waveguide sandwiched between two thin films with a low refractive index, which act as potential barriers. The theory of the FTIR effect is presented in [2, 3]. In [4] a circular polarization beam splitter based on FTIR effect was considered. High performance FTIR based thin-film linear polarizing beam splitter has been demonstrated in [5]. It is well known that FTIR structures have transmission peaks at certain wavelengths and angles, which are very sensitive to the parameters of prisms and embedded layers. It was shown in [6] that the divergence of the incident light beam when measuring the spectral characteristics of FTIR filters should not exceed a certain limit value, which does not exceed several angular minutes. Spatial-frequency filtering is widely used to improve the image in color visualization systems, color display devices, etc. In [7], a device was proposed for the spatial separation of an incident white light beam into three color beams. In [8] the effect of frequency dispersion in the resonator layer on the FTIR process has been considered. The FTIR effect of nanoparticles embedded in the resonator layer of the filter on its transmission spectrum was investigated theoretically [8]. It was shown that the incident beam of a given wavelength is split into three angularly separated beams. Splitting of the filter bandwidth into three narrowband spectral lines for the specific incidence angle was demonstrated.

In this paper, the effect of anomalous frequency dispersion caused by silver and gold nanoparticles embedded in the central and lateral layers of the FTIR filter on the resonant transmission of light is theoretically investigated. In contrast to the system considered in [8], here we study the effect of plasmonic nanoparticles embedded in both the central and lateral layers. This allows us to obtain five resonant bands at once for a given angle of incidence.

## 2. FTIR filter

A schematic model of the filter with nanoparticle inclusions in the layers is shown in Fig. 1. The device consists of a three-layer structure placed between two prisms with refractive indices $n_p$ and $n'_p$. The three-layer structure consists of a central layer with a high refractive index and thickness $d_2$, sandwiched between two films with low refractive indices $n_1$ and $n'_1$, and thicknesses $d_1$ and $d'_1$, respectively. Chromatic and angular filtering of the incident light beam occurs due to the resonant diffraction effect, when light propagates through an inhomogeneous stratified medium (layered structure), i.e., the FTIR effect.

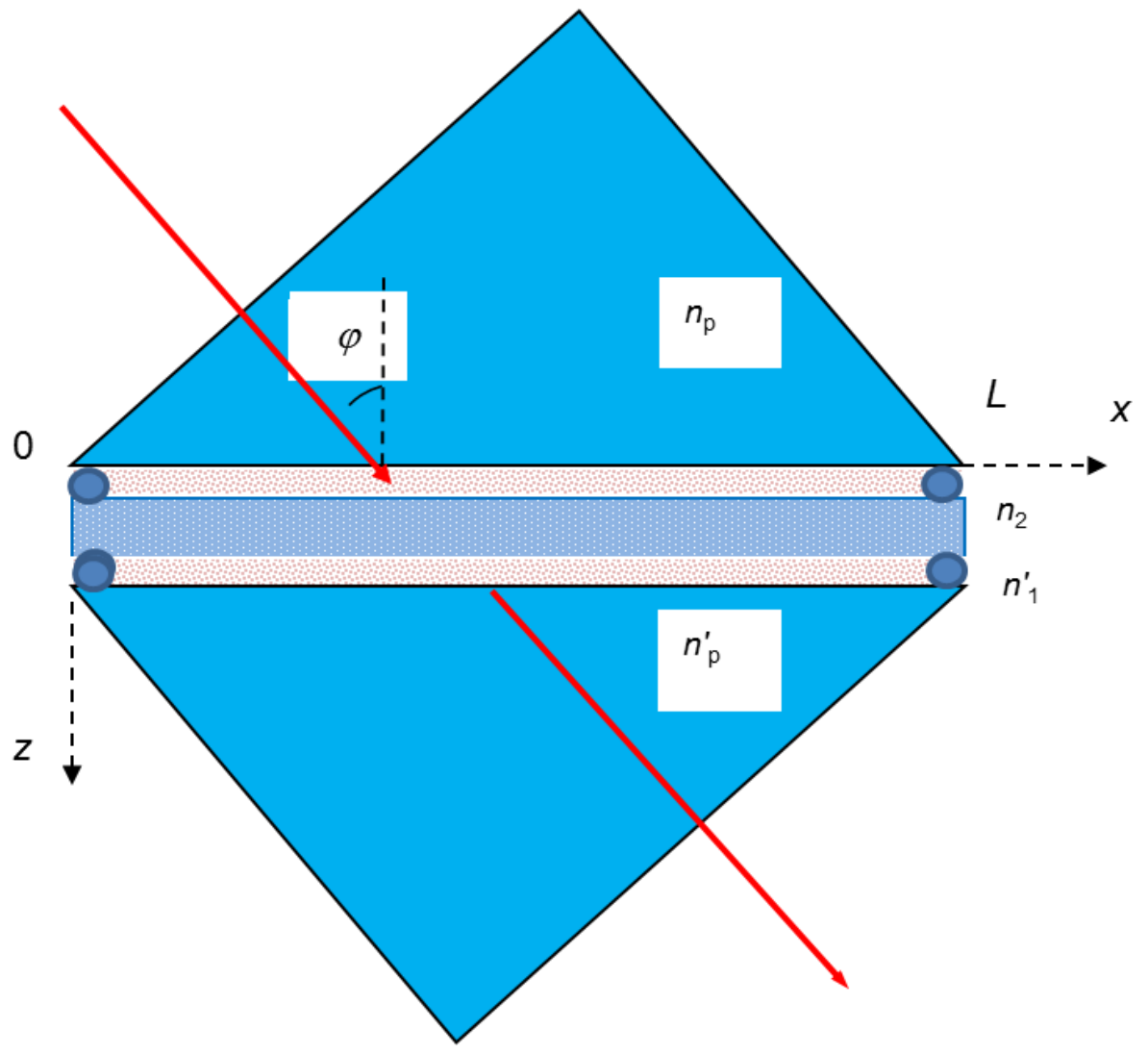

**Figure 1**. FTIR filter with nanoparticle inclusions in the central layer and side layers.

*2.1 Resonance condition*

For the s-polarization of incident light, the resonant condition that follows from the solution of Maxwell's equations, taking into account the boundary conditions, is determined by the equation [7, 8]:

$$k'_z d_2 = -\,arctan\left[\frac{k'_z(q_z+q'_z)}{q_z q'_z - {k'}_z^2}\right] - Q_3, \tag{1}$$

where $Q_3 = 2q_z k'_z(q_z^2 - k_z^2)\,exp(-2q_z d_1)/[(q_z^2 + k_z^2)(q_z^2 + {k'}_z^2)$, $q_z = (\omega/c)\sqrt{n_p^2 \sin^2\varphi - n_1^2}$ , $k_z = (\omega/c)\sqrt{n_p^2 - n_p^2\, sin^2\,\varphi}$, $q'_z = (\omega/c)\sqrt{n_p'^2 \sin^2\varphi - n_1'^2}$, $k'_z = (\omega/c)\sqrt{n_2^2 - n_p'^2\, sin^2\,\varphi}$ are the wavenumbers, $\omega$ is the frequency, $c$ is the speed of light, $\varphi$ is the incident angle of a beam.

Similar expression for the resonance condition can be obtained in the case of *p*- polarization.

*2.2. Influence of Frequency Dispersion*

The dependence of the angle of incidence on the dispersion follows from the resonant condition (1):

$$\varphi(\omega) = \varphi_0(\omega) + \frac{d_2^*}{d_{2eff}}\frac{\Delta\varepsilon'_2}{n_p^2\, sin\, 2\varphi_0} + \left(1 - \frac{d_2^*}{d_{2eff}}\right)\frac{2\Delta\varepsilon'_1}{n_p^2\, sin\, 2\varphi_0}, \tag{2}$$

where $\Delta\varepsilon'_2$ and $\Delta\varepsilon'_1$ are the changes of the real part of the permittivity in the central and side layers, respectively, $\varphi_0(\omega)$ is the incident angle in the absence of dispersion, $d_2^* = d_2 + 2q_z/\left(q_z^2 + {k'}_z^2\right)$, $d_{2eff} = d_2 + 1/q_z + 1/q'_z$.

Note that the second term in equation (2) takes into account the dispersion in the central layer, and the third term takes into account the dispersion in the side layers of the filter.

*2.2.1. Plasmon resonance in metal nanoparticles*

There are two types plasmon modes: propagating surface plasmons and localized surface plasmons. Below we consider the localized plasmon resonances associated with metallic nanoparticles. Localized surface plasmons are combined oscillation of free electrons in a metallic nanoparticle and associated oscillations of the electromagnetic field. The resonance frequency depends on the size, shape, and local optical environment of the particle [9, 10] and usually occurs in the visible to near-infrared part of the spectrum for noble metal (Ag, Au) nanostructures.

Below we will consider filter layers with embedded metal nanoparticles. The size of the nanoparticles is assumed to be significantly smaller than the wavelength of light and they are randomly distributed. Such a medium in the framework of the Maxwell-Garnett model is described by the effective permittivity, which for spherical nanoparticles satisfies the equation [11, 12]:

$$\varepsilon_{eff} = \varepsilon_m + \frac{3\eta(\varepsilon_p-\varepsilon_m)\varepsilon_m}{3\varepsilon_m+(1-\eta)(\varepsilon_p-\varepsilon_m)}, \tag{3}$$

where $\eta$ is the volume fraction of nanoparticles, $\varepsilon_m$ is the dielectric permittivity of the central layer, and $\varepsilon_p$ is the dielectric permittivity of the nanoparticles.

The Maxwell–Garnett model is in good agreement with experimental data if particles are smaller than the wavelength of radiation and the distances between particles are larger than their radii. An advantage of such model is that the propagation of radiation in a heterogeneous medium can be analysed without solution of Maxwell's equations at each point of space and disregarding scattering on particles composing the heterogeneous medium and interference of scattered waves.

In the framework of the Drude model the optical properties of metal nanoparticles will be described by the expression:

$$\varepsilon_p(\omega) = \varepsilon_0 - \frac{\omega_p^2}{\omega^2+i\omega\gamma}, \tag{4}$$

where $\varepsilon_0$ is the parameter describing the contribution of the bound electrons to the polarizability, $\omega_p$ is the plasma frequency of a free electron gas, and $\gamma$ is the damping factor of plasma oscillations. For metal nanoparticles the damping factor $\gamma$ is a size-dependent function [10, 13, 14]:

$$\gamma(a) = \gamma_0 + q\frac{v_F}{a}, \tag{5}$$

where $v_F$ is the velocity of electrons at the Fermi energy, $\gamma_0$ is the damping factor for an unlimited volume of metal, $a$ is the radius of the nanoparticle. The factor $q$ is determined by the processes of electron scattering on the surface of nanoparticles, and it is usually assumed to be equal to unity.

Figure 2 shows the dependences of the real and imaginary parts of the value $\Delta\varepsilon_2 = \varepsilon_{eff} - \varepsilon_m$ on the wavelength. Note that plasmon resonances occur in the visible spectral range.

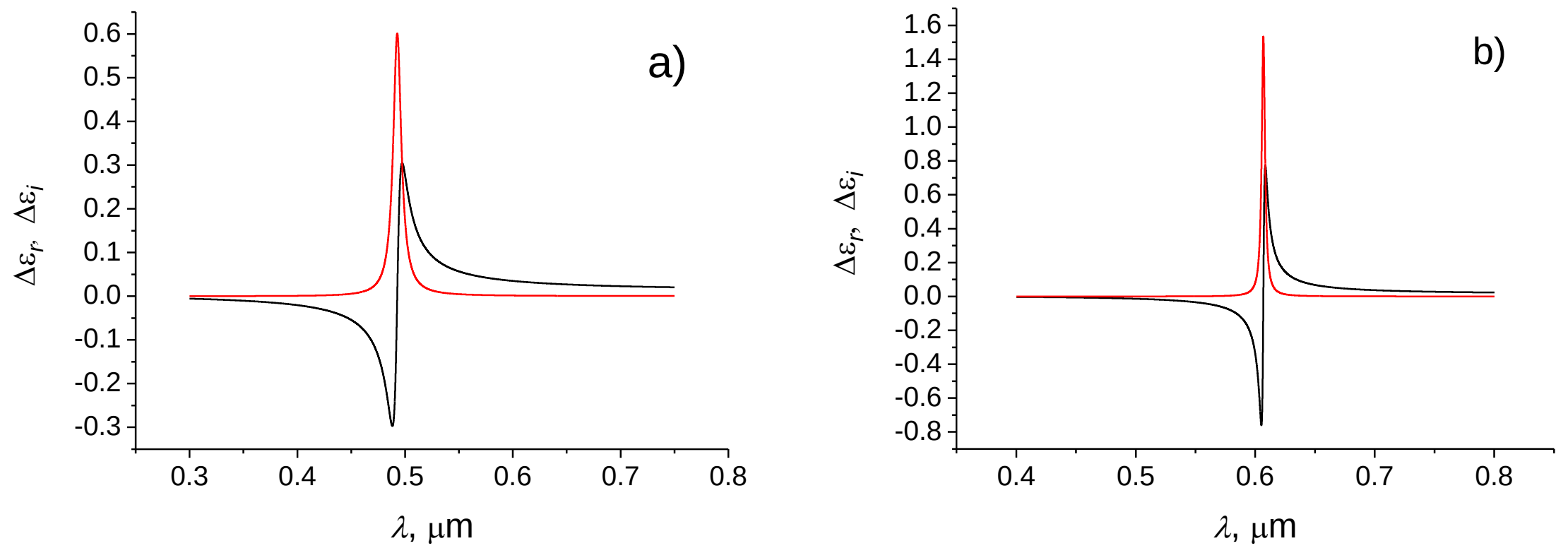

**Figure 2.** Dispersion $\Delta\varepsilon_r$ (black color) and absorption $\Delta\varepsilon_i$ (red color) as function of wavelength for Ag (a) and Au (b) nanoparticles. $a$ = 30 nm, $\eta = 10^{-3}$.

Parameters for silver nanoparticles: $\varepsilon_0 = 5$, $\omega_p = 1.38 \cdot 10^{16}$ s$^{-1}$, $\gamma_0 = 2.39 \cdot 10^{13}$ s$^{-1}$ [13,15] and $v_F = 1.4 \cdot 10^6$ m s$^{-1}$ [14]. The plasmon resonance in a medium with $\varepsilon_m = 4.0$ occurs at a wavelength $\lambda_p \simeq 493$ nm. For a gold nanoparticle with parameters $\varepsilon_0 = 9.5$, $\omega_p = 1.3 \cdot 10^{16}$ s$^{-1}$, $\gamma_0 = 1.67 \cdot 10^{13}$ s$^{-1}$ [13, 15], the plasmon resonance occurs at a wavelength $\lambda_p \simeq 607$ nm.

## 3. Simulation results

Let's now consider the effect of frequency dispersion caused by silver and gold nanoparticles embedded in the central and lateral layers of the FTIR filter on the resonant transmission of light.

Figure 3 shows the dependence of the resonant angle of incidence on the wavelength of the *s*-polarized beam using the parameters for silver and gold in the Drude model. It is seen that a change in the angle of incidence leads to a shift of the resonant spectral lines. It is possible to get several resolved spectral lines at the output by changing the angle of incidence. It follows from the simulation that the frequency dispersion of a heterogeneous layer causes angular splitting of the beam for a given wavelength of incident light. When the wavelength of the incident beam changes, up to five spatially separated beams can be observed at the output.

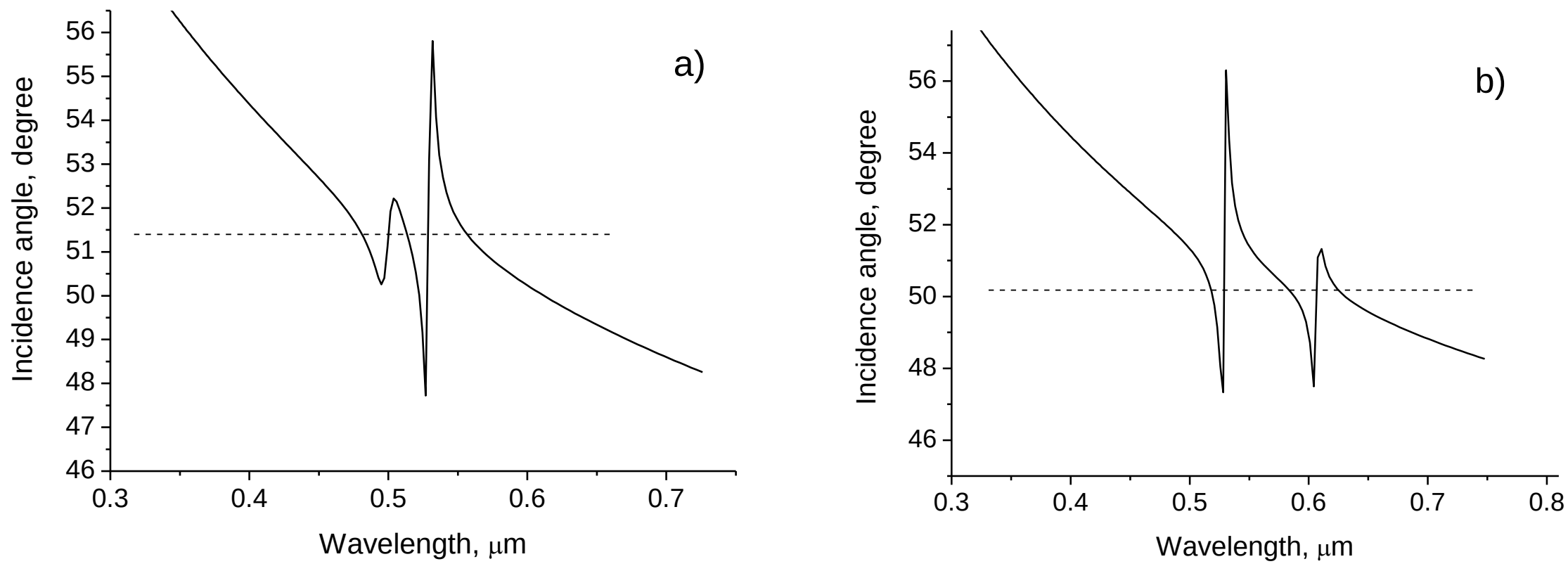

**Figure 3.** Dependence of resonance incident angle of *s*-polarized beam on the wavelength. $n_p$ = 2.0, $n_1$ = 1.38, $n_2$ = 2.0, $d_2$ = 70 nm; $\eta = 10^{-3}$. (a) $d_1$ = 500 nm; Ag nanoparticles in the central layer and Au nanoparticles in the side layers; (b) $d_1$ = 300 nm; Au nanoparticles in the central and side layers.

In Fig. 4 the spectral shapes of *s*- polarized transmitted light are shown for the central and lateral layers with gold nanoparticles. The spectral bandwidths decrease with the increase of the low-index layer thickness $d_1$ and the spectral line width $\Delta\lambda = 27$ nm was obtained at $\lambda$ = 518 nm and $d_1$= 300 nm.

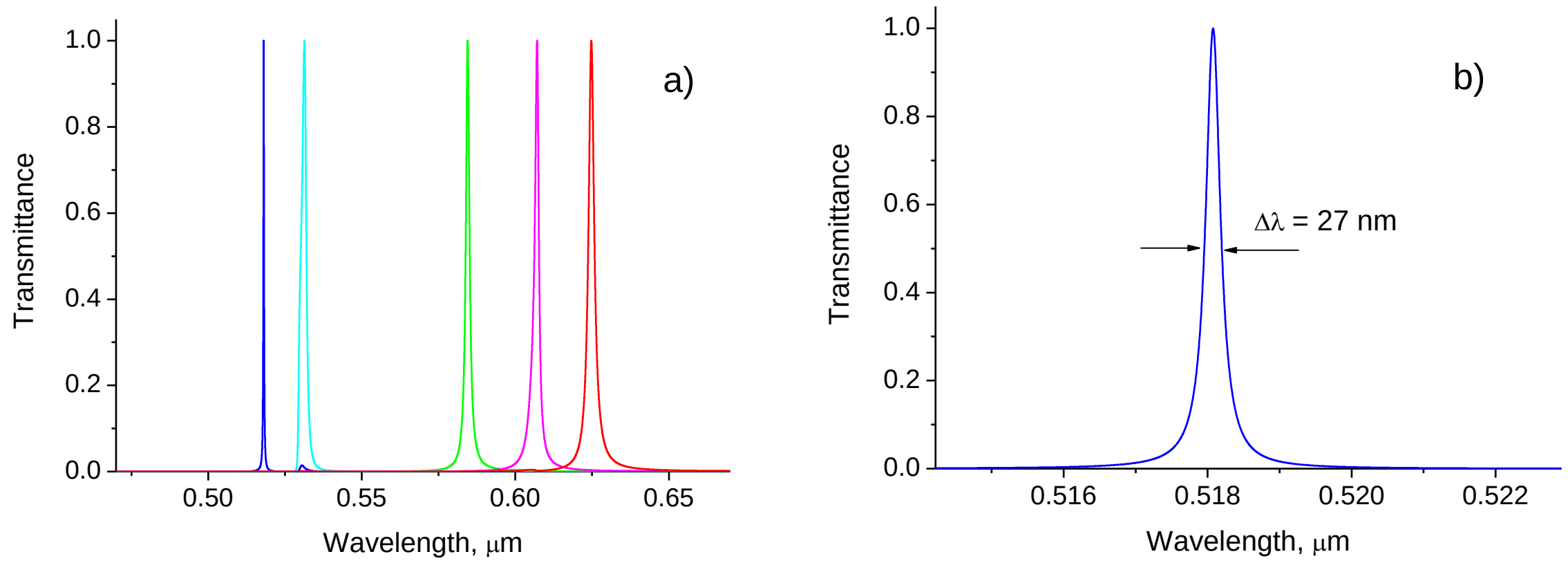

**Figure 4.** Spectral lines of transmitted light: $n_p$ = 2.0, $n_1$ = 1.38, $n_2$ = 2.0, $d_1$ = 300 nm, $d_2$ = 70 nm, $\eta = 10^{-3}$, $\varphi = 50.18°$.

Figure 5 shows the dependence of the resonant angle of incidence on the wavelength of the *s*-polarized beam for embedded *Ag* nanoparticles in the central layer and *Au* nanoparticles in the lateral layers.

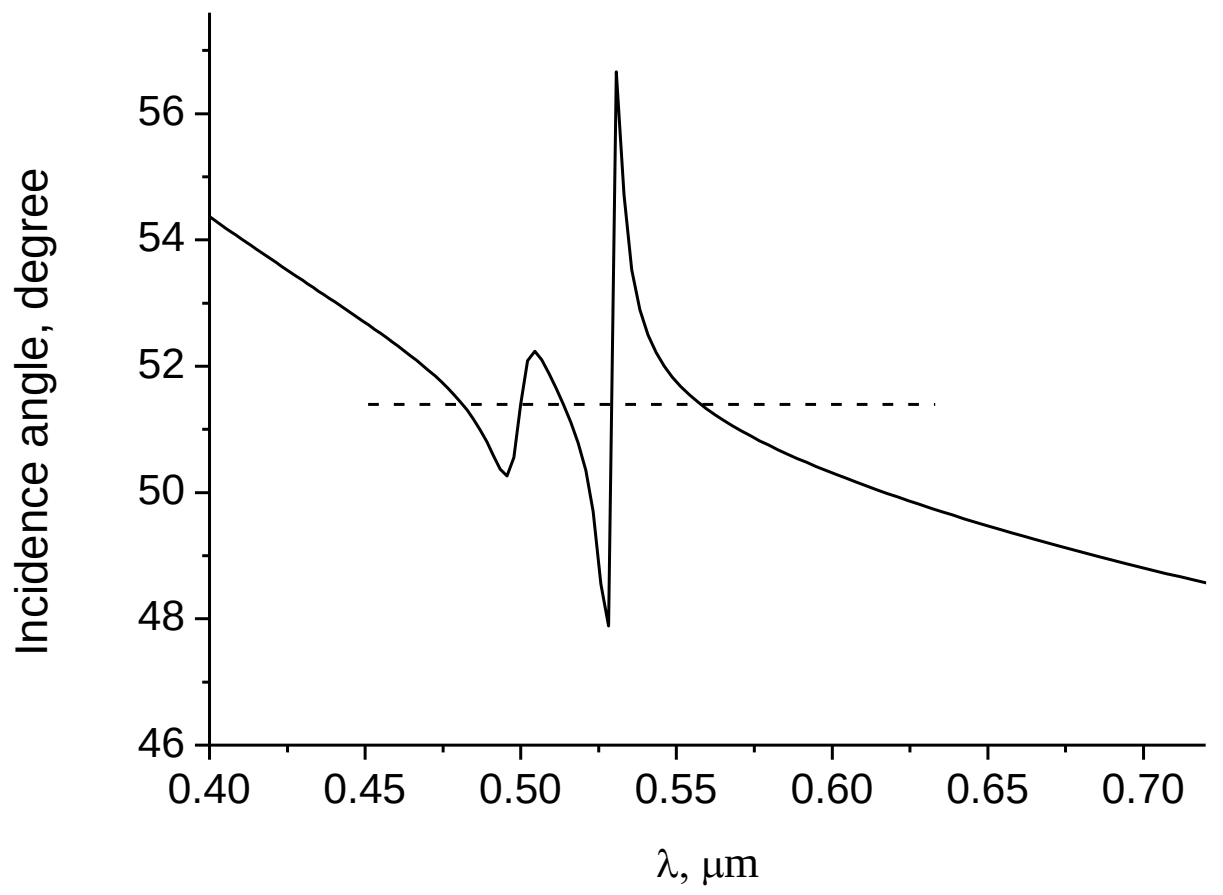

**Figure 5.** Dependence of resonance incident angle of *s*-polarized beam on the wavelength. $n_p$ = 2.0, $n_1$ = 1.38, $n_2$ = 2.0, $d_2$ = 70 nm; $\eta = 10^{-3}$, $d_1$ = 600 nm; Ag nanoparticles in the central layer and Au nanoparticles in the side layers.

In Fig. 6 the spectral lines of *s*- polarized transmitted light corresponding to the incident angle $\varphi = 51.49^{\circ}$ are shown for silver nanoparticles embedded into a central layer and gold nanoparticles embedded into the lateral layers.

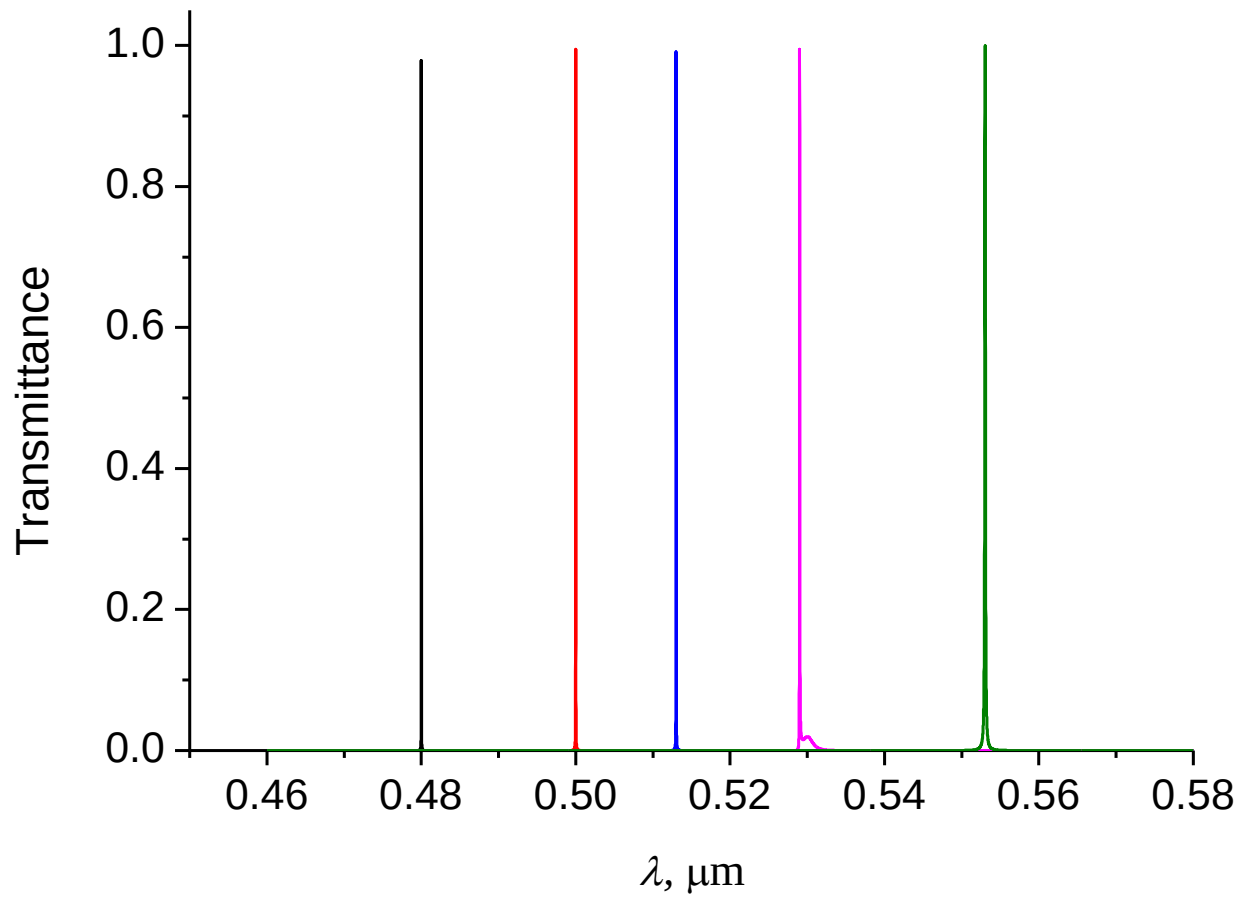

**Figure 6.** Spectral lines of transmitted light: $n_p$ = 2.0, $n_1$ = 1.38, $n_2$ = 2.0, $d_1$ = 600 nm, $d_2$ = 70 nm, $\eta = 10^{-3}$, $\varphi = 51.49^{\circ}$.

It is seen that for a given angle of incidence, there are five resonant bands at once. This indicates that the resonance condition in the resonator is satisfied for five wavelengths simultaneously.

In Fig. 7 the spectral lines corresponding to the resonant wavelengths $\lambda = 0.48\ \mu m$ (Fig.7a) and $\lambda = 0.553\ \mu m$ (Fig.7b) are presented.

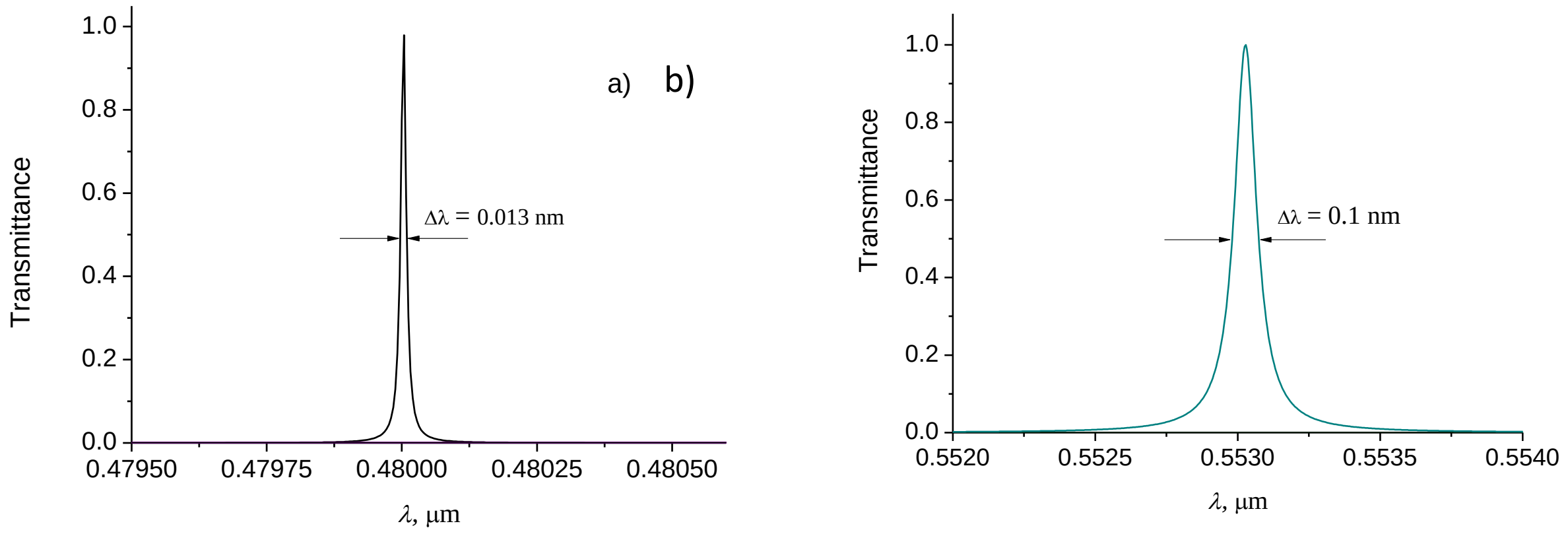

**Figure 7.** Spectral lines of transmitted light from Fig. 6 on an enlarged scale.

The width of the spectral bands decreases with decreasing wavelength, and the spectral lines with a width of $\Delta\lambda = 0.013$ nm and $\Delta\lambda = 0.1$ nm were obtained at $d_1$ = 600 nm for the central wavelengths $\lambda = 0.48\ \mu m$ and $\lambda = 0.553\ \mu m$, respectively.

It can be seen that there are five resonant bands at once for a given angle of incidence. This indicates that the resonance condition in the resonator is satisfied for five wavelengths simultaneously. This property of the device can be applied in visualization systems [16]. It should be noted that acousto-optic filters with inclusions of dielectric nanoparticles, considered in [12], also provide ultra-narrow spectral lines due to resonant Bragg diffraction on a periodic volume grating created by ultrasound in a crystal.

## 4. Discussion and conclusions

It is known that resonant phenomena can arise for wave propagation in an inhomogeneous plane-layered media, which lead to a sharp increase in the transmission of waves with a certain wavelength. In quantum mechanics a similar effect is seen when the resonant transmission of de Broglie waves through a system of two potential barriers (the Ramsauer effect) occurs. Such phenomena take place in the propagation of waves in plasma, in the optical systems of the type of interference filters, Fabry-Perot resonators [17, 18], etc. Recently, a refractive index sensor in the terahertz domain comprising a one-dimensional photonic bandgap structure with plasmonic inclusions was proposed for optofluidics [19].

Resonant structures based on heterogeneous inclusions open up new opportunities for the creation of infrared (IR) and terahertz technology devices that are inaccessible to conventional materials. In [20], resonant tunneling and the Goos–Hänchen shift were investigated in the FTIR configuration coated by a graphene sheet. Tunable resonance Goos–Hänchen and Imbert-Fedorov shifts for terahertz beams reflected from graphene plasmon metasurfaces were studied in [21]. In [22], large Goos–Hänchen shifts were demonstrated near the surface plasmon resonance in subwavelength gratings.

Future research may be related to solving the problem of light diffraction in the FTIR filter, taking into account the spatial limitations of the incident light beam. Of particular interest is the consideration of structured light beams with orbital angular momentum [23, 24]. It is also of practical interest to study the effects of the spin-orbit interaction in a FTIR filter that occur when considering two-dimensional vortex incident beams. This will allow to introduce an additional parameter for controlling the output data of the radiation beam, which can be used in the development of new apparatus and devices. Wide possibilities open up when using inclusions of nanoparticles with a complex shape and with special magnetic, dielectric and conductive properties.

Thus, a new FTIR filter is proposed, based on a combination of photonic and plasmonic effects, which lead to a narrowing of the bandwidth and a transformation of the transmission spectrum. It is shown that an incident beam of a given wavelength splits into set of beams with angular separation. The splitting of the filter bandwidth into several narrow-band spectral lines for a given angle of incidence is shown. The splitting of the filter bandwidth into five narrow-band spectral lines for a given angle of incidence is shown for embedded silver and gold nanoparticles. This type of thin-film filter can be useful in many applications, including spectroscopy, sensors, in spectral regions extending from the UV to the far IR range, as well as in color imaging systems.

**Acknowledgements**

This research was funded by the Ministry of Science and Higher Education of the Russian Federation under the State contract FFNS-2022-0009 and by the Russian Foundation for Basic Research, project number 19-29-11026.